\documentclass[useAMS,usenatbib]{mn2e}

\usepackage[T1]{fontenc}
\usepackage{aecompl}

\usepackage{verbatim}
\usepackage{graphicx}

\makeatletter

\newcommand{\apj}{ApJ}
\newcommand{\apjl}{ApJ}
\newcommand{\aj}{AJ}
\newcommand{\mnras}{MNRAS}
\newcommand{\aap}{A\&A}
\newcommand{\aaps}{A\&AS}
\newcommand{\araa}{ARA\&A}
\newcommand{\apjs}{ApJS}
\newcommand{\procspie}{Proc. SPIE}

\title[A Transition Mass in the Local Tully-Fisher Relation]{A Transition Mass in the Local Tully-Fisher Relation}
\author[Simons et al.]{Raymond C. Simons$^{1}$\thanks{E-mail: rsimons@jhu.edu}, Susan A. Kassin$^{2}$, Benjamin J. Weiner$^{3}$, Timothy M. Heckman$^{1}$,
 \newauthor Janice C. Lee$^{2}$, Jennifer M. Lotz$^{2}$, Michael Peth$^{1}$ and Kirill Tchernyshyov$^{1}$\\\\
$^{1}$Johns Hopkins University, Baltimore, MD\\
$^{2}$Space Telescope Science Institute, 3700 San Martin Drive, Baltimore, MD\\
$^{3}$Steward Observatory, University of Arizona, Tucson, AZ}

\begin{document}
\date{Accepted ??. Received ??; in original form ??}

\pagerange{\pageref{firstpage}--\pageref{lastpage}} \pubyear{2014}

\maketitle

\label{firstpage}

\begin{abstract}

We study the stellar mass Tully-Fisher relation (TFR; stellar mass versus rotation velocity) for a morphologically blind selection of emission line galaxies in the field at redshifts 0.1 $<$ z $<$ 0.375. Kinematics ($\sigma_g$, V$_{rot}$) are measured from emission lines in Keck/DEIMOS spectra and quantitative morphology is measured from V- and I-band Hubble images. We find a transition stellar mass in the TFR, $\log$ M$_*$ = 9.5 M$_{\odot}$.  Above this mass, nearly all galaxies are rotation-dominated, on average more morphologically disk-like according to quantitative morphology, and lie on a relatively tight TFR. Below this mass, the TFR has significant scatter to low rotation velocity and galaxies can either be rotation-dominated disks on
the TFR or asymmetric or compact galaxies which scatter off.  We refer to this transition mass as the ``mass of disk formation", M$_{\mathrm{df}}$ because above it all star-forming galaxies form disks (except for a small number of major mergers and highly star-forming systems), whereas below it a galaxy may or may not form a disk.
\end{abstract}


\section{Introduction}

The Tully-Fisher relation (TFR; \citealt{1977A&A....54..661T}) is an empirical scaling between the luminosity of a disk galaxy and its rotation velocity. Later work replaced luminosity with the more physical quantity stellar mass, M$_*$. In the local Universe, the TFR is remarkably tight \citep{ 2001ApJ...563..694V, 2001ApJ...550..212B, 2005ApJ...633..844P, 2006ApJ...643..804K, 2007ApJ...671..203C, 2008AJ....135.1738M, 2011MNRAS.417.2347R} and its parameterization serves as an important constraint for models of disk galaxy formation (e.g. \citealt{1998MNRAS.295..319M, 1999MNRAS.310.1087S, 2000ApJ...538..477N,2007ApJ...654...27D}).

Although the TFR is well behaved for ordered disk galaxies, morphologically disturbed or compact galaxies tend to fall below the local relation, exhibiting low rotation velocity for a given stellar mass \citep{2007ApJ...660L..35K, 2012ApJ...758..106K}. Ordered velocity fields may be disturbed in major merging events or tidal disruptions \citep{2005MNRAS.356.1177R, 2007A&A...473..761K, 2010ApJ...710..279C, 2012A&A...546A..52D}, through the accretion of non-ordered external angular momentum from cold flows \citep{2009ApJ...694..396B, 2010ApJ...712..294E} and/or disruptive feedback \citep{1999ApJ...513..142M, 2009ApJ...699.1660L}.

\citet{2006ApJ...653.1049W} and \citet{2007ApJ...660L..35K} demonstrated that accounting for both the disordered motions ($\sigma_g$) and ordered rotation velocity (V$_{rot}$) in a new kinematic quantity,  $S_{0.5} = \sqrt{0.5V_{rot}^2+\sigma_g^2}$, re-establishes a tighter $S_{0.5}$-M$_*$ scaling relation. This relation is independent of galaxy morphology and coincident with the Faber-Jackson relation for early type galaxies. Numerical simulations have shown that S$_{0.5}$  traces the overall potential well of galaxy-dark halo systems, even for galaxies undergoing drastic kinematic events, such as a major merger \citep{2010ApJ...710..279C}. The measurement of a disordered velocity component has been incorporated more in recent years (e.g. \citealt{2009ApJ...706.1364F, 2010A&A...510A..68P, 2010MNRAS.402.2291L, 2012MNRAS.420.1959C, 2012A&A...546A.118V, 2014ApJ...795L..37C, 2015ApJ...799..209W}).

With further investigation, \citet{2012ApJ...758..106K} demonstrated that since z $\sim$ 1 star-forming galaxies have been kinematically settling from systems with a dominant dispersion component to rotationally supported disks. Such galaxies follow kinematic downsizing: more massive galaxies exhibit the most ordered kinematics (high V$_{rot}$/$\sigma_g$) at all epochs. While high and intermediate mass galaxies are likely settling to disks, it is unclear whether low mass galaxies follow the same evolutionary path.

It is well known locally that morphological type is a strong function of stellar mass, with dwarf systems exhibiting more irregular morphology, distinct from their large disk counterparts (e.g. \citealt{1994ARA&A..32..115R, 2009MNRAS.400..154B, 2015MNRAS.446.2967M}). Moreover, there is mounting evidence that disturbed galaxies are increasingly more common at low masses in the early Universe \citep{ 2013MNRAS.433.1185M}.

However, kinematic surveys often select against galaxies with disturbed morphologies with regards to other interests, e.g. performing distance measurements or studying dark matter (e.g. \citealt{2010ApJ...716..198B}). In the local Universe, dwarf galaxies can show rotational signatures in both their HI and stellar components (e.g. \citealt{2002A&A...390..829S, 2009A&A...493..871S, 2012AJ....144....4M}). When the morphological selection is opened to include irregular galaxies, compact galaxies, and close pairs, the presence of features such as peculiar velocity fields and thick disks are found. Indeed, complex kinematics and low rotational support is frequently found for both low luminosity  (e.g. \citealt{2001AJ....121..625B, 2002AJ....123.2358K, 2003ApJ...592..111Y, 2005AJ....130.1593V,2014MNRAS.439.1015K, 2014ApJ...795L..37C}) and highly star-forming dwarfs (e.g. \citealt{1998AJ....116.1186V, 2004ApJ...607..274C, 2014A&A...566A..71L}). However, locally, limited studies have been performed placing large samples of these disordered systems on the TFR, or relating the relative contributions of V and $\sigma$ to the morphology of the galaxy.

In this paper we study the resolved kinematics for a morphologically unbiased sample of star-forming galaxies at z $\sim$ 0.2. Compared to nearby samples, a survey at z $\sim$ 0.2 benefits from the smaller angular sizes of galaxies and the higher target density. Large homogenous samples can be efficiently obtained with a multi-object spectrograph. This sample is unique from other local samples in that we include galaxies with both disturbed and disk-like morphologies. Furthermore, we incorporate measurements of the non-negligible contributions from random motions as well as the rotation velocity.

In section 2, we discuss the sample selection, the method used for measuring kinematics from emission lines and the measurements of quantitative morphologies. In section 3, we present the main result of this paper: a transition stellar mass in the TFR. In section 4, we illustrate the correlation between gas phase kinematics and galaxy morphology. In section 5, we compare our results with measurements of kinematics for low mass galaxies in the literature both locally and at intermediate redshift. Our conclusions are presented in section 6. In the Appendix, we test our ability to recover kinematics for the galaxies with the smallest angular extent in our sample. We adopt a $\Lambda$CDM cosmology defined with (h, $\Omega_m$, $\Omega_{\Lambda}$) = (0.7, 0.3, 0.7).

\section{Data and Sample Selection}
In this section we detail the sample selection and the measurements of the kinematics and morphological indices. We reference the Appendix for further analysis and discussion of the effects of seeing on our kinematic measurements.

We focus on the lowest redshift bin in the galaxy sample used by \citet{2007ApJ...660L..35K} and \citet{2012ApJ...758..106K} (hereafter referred to as K07 and K12, respectively), namely 0.1 $<$ z $<$ 0.375. At these redshifts, the K12 sample is sensitive to low mass dwarf galaxies. We briefly review how this sample was selected and defer to K12 for further details. The K12 sample is drawn from field 1 of the DEEP2 Redshift Survey \citep{2013ApJS..208....5N}. DEEP2 employed the DEIMOS multi-object spectrograph \citep{2003SPIE.4841.1657F} on the Keck-II telescope. The 1200 line/mm grating was used and the slits were fixed at 1$\arcsec$, leading to a spectral resolution of R $\sim$ 5000. The K12 sample was cut on nebular line strength ($>10^{-17} $ erg s$^{-1}$ cm $^2$), available Hubble Space Telescope (HST) imaging and slit alignment from the photometric major axis ($\le$ 40$^{\circ}$).

We use HST imaging in two passbands with the Advanced Camera for Surveys (ACS), V (F606W) and I (F814W), from the AEGIS survey \citep{2007ApJ...660L...1D}. Each of the images has a pixel scale of 0.03$\arcsec$ and a typical FWHM PSF of 0.1$\arcsec$ (0.33 kpc at z = 0.2). Inclinations are measured with the SExtractor software \citep{1996A&AS..117..393B} using the V-band HST image. The V-band traces the young stars and thus should not be very different from the nebular emission line morphology. A further cut is applied to the sample, limiting inclinations to 30$^{\circ} <$ i $< 70^{\circ}$, to avoid uncertain inclination corrections for face-on galaxies and dust effects in edge on systems. However, a handful (4) of severely disturbed galaxies for which inclinations and PAs are uncertain are included. The exclusion of these galaxies does not alter our conclusions.

Stellar masses were derived using the rest frame B-V color and absolute B-band magnitude \citep{2001ApJ...550..212B, 2005ApJ...625...23B} with refined empirical corrections from SED fitting \citep{2006ApJ...651..120B}, as described in \citet{2007ApJ...660L..51L}. The adopted IMF is from \citet{2003PASP..115..763C}. Errors on stellar masses are approximately 0.2 dex.

Spectral slits must be aligned to within 40$^\circ$ of a galaxy's major axis to reliably recover rotation (\citealt{2006ApJ...653.1027W}). K07 removed galaxies with slits misaligned by more than 40$^\circ$ from the HST derived photometric major axis, except for the 4 galaxies with severely disturbed morphologies. Beyond this cut there are no residual correlations between measured kinematics (rotation or dispersion) and slit alignment in our sample. The scale of the atmospheric turbulence sets a limiting angular size for measuring rotation velocity in a galaxy. The seeing tends to smooth and eliminate small scale rotation gradients. We test the limiting resolving power imposed by beam smearing in the Appendix. For the observational conditions of DEEP2 and for the lowest S/N in our sample, kinematics can be measured for galaxies to diameters encompassing 95$\%$ of the light (D$_{95}$) of (0.87$\pm$0.06) $\times$ seeing. We make a further cut on the K07 sample and remove 12 galaxies which are not extended enough to confidently measure kinematics.

Our final sample contains 119 galaxies and uniformly covers the ``blue cloud" (Figure 1 in K12). The qualifying feature of our selection is that, aside from the inclination cut, there was no explicit selection on morphology. Contrary to previous TFR studies, this sample selection includes disturbed and compact galaxies. This allows us to more fully sample the population of emission line galaxies.

\subsection{Kinematics}

We adopt kinematics measured from bright nebular emission lines (H$\alpha$ $\lambda$6563 or [OIII] $\lambda$5007) in K07. The rotation velocity, uncorrected for inclination, and spatially resolved gas dispersion, $\sigma_g$, were measured with the program ROTCURVE \citep{2006ApJ...653.1027W}, taking into account the effects of seeing. The seeing was measured through alignment stars on each slit mask (see \citealt{2013ApJS..208....5N}). For our sample, the seeing varied between 0.55$\arcsec$ and 1.2$\arcsec$ FWHM, with a median value of 0.75$\arcsec$ (1.36 kpc at z = 0.1, 3.84 kpc at z = 0.375).

The kinematic measurements have been described in previous papers, so we will briefly review them here and refer to \citet{2006ApJ...653.1027W} for the details. In short, ROTCURVE builds model arctan rotation curves where V is the velocity on the flat part of the rotation curve (V$_{rot}$ $\times$ $\sin (i)$), with an additional dispersion term ($\sigma_g$), which is constant with radius. Due to the seeing, the rise of the rotation curve is not well resolved and the turnover radius for the rotation curve (r$_t$) is not well constrained. The turnover radius is kept fixed to a value of 0.2$\arcsec$. The recovered rotation velocity and turnover radius are only slightly covariant, with a marginal $\pm$ 0.1 dex change in V for a $\pm$0.1$\arcsec$ change in r$_t$. Models with varying V  and $\sigma_g$ are blurred with the seeing and fit to the data. V and $\sigma_g$ are explored with a grid spacing of 5 km s$^{-1}$ and the best fit value is determined through a $\chi^2$ minimization on this grid.

The spectral resolution of the 1200 line/mm grating used by the DEEP2 survey allows measurements for V$_{rot}$ $\times$ $\sin (i)$ down to $\sim$ 5 km\,s$^{-1}$ and $\sigma_g$ to $\sim$ 15 km\,s$^{-1}$. Measurements with best fits below these limitations are set to these values as upper limits.  This is the case for 7 galaxies for V$_{rot}$ and 22 galaxies for $\sigma_g$. Example kinematic fits and the corresponding $\chi^2$ spaces are shown later in this paper in Figure \ref{fig:indmorph}.

\begin{figure}
\begin{centering}
\includegraphics[angle=0,scale=.37]{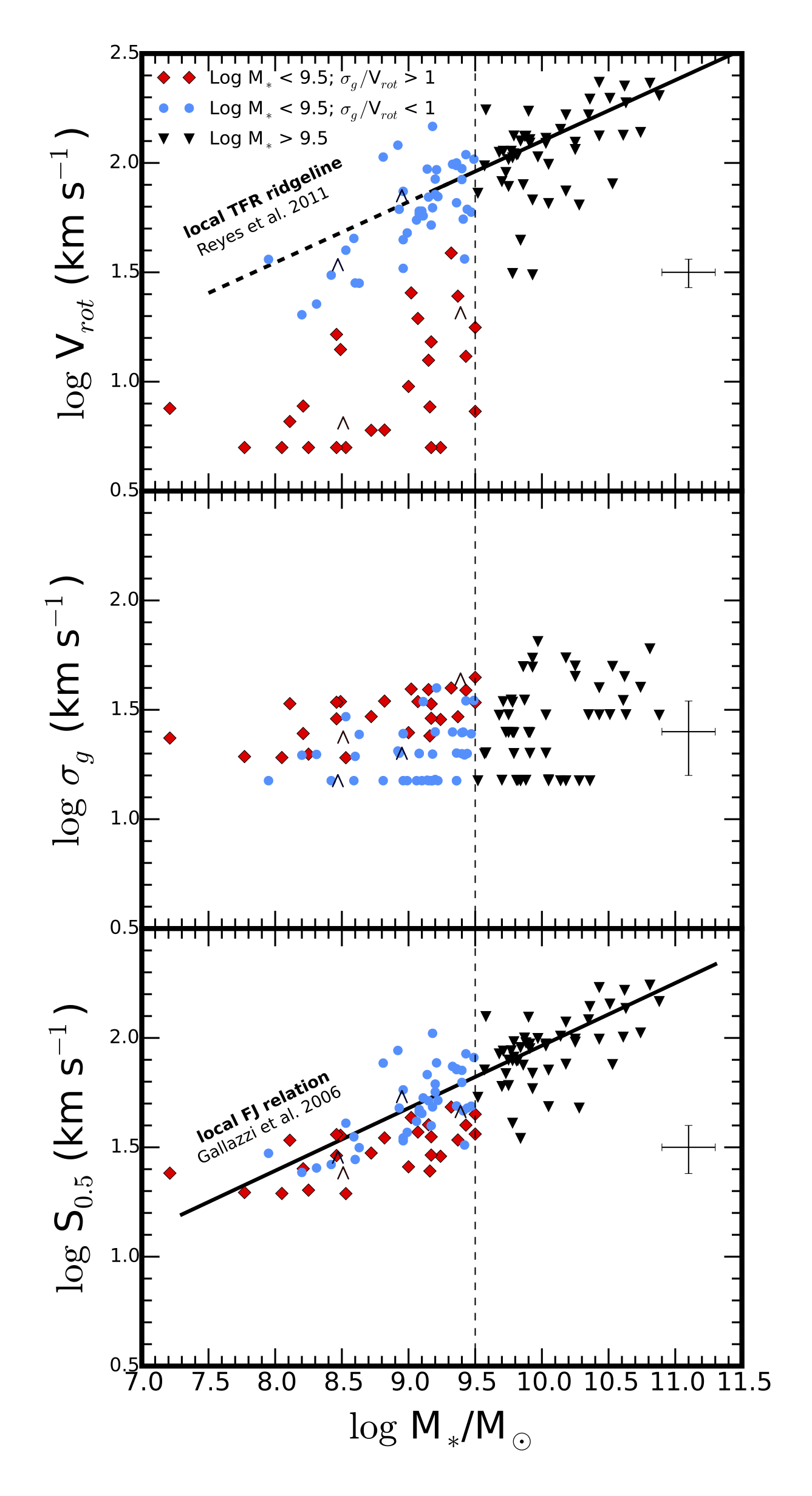}
\caption {(a) The stellar mass Tully-Fisher relation for the K07/K12 sample of galaxies at 0.1 $<$ z $<$ 0.375 is shown.  Below a stellar mass of $\log$ M$_*$/M$_{\odot}\cong$ 9.5, there is significant scatter to low V$_{rot}$ (red points). The local TFR \citep{2011MNRAS.417.2347R} (black line), adjusted for a \citet{2003PASP..115..763C} IMF, and its extrapolation to lower stellar masses (dashed line) is shown. b) For the same galaxies the integrated gas velocity dispersion, a measure of disordered motions, is shown versus stellar mass. c) Accounting for both ordered and disordered motions in the S$_{0.5}$-M$_*$ relation results in a relatively tight scaling, coincident with the local Faber-Jackson relation from \citet{2006MNRAS.370.1106G}. Galaxies are color/shape coded according to whether they are on the TF ridgeline and whether they have high or low stellar mass. This coding is repeated in other plots in the paper. The rotation velocities for the four galaxies with carats are not inclination corrected due to their highly uncertain inclination.}
\label{fig:kinematics}
\end{centering}
\end{figure}

HST imaging is used to measure inclinations in order to correct the measured rotation velocity. As mentioned previously, corrections were not applied to four galaxies with severely disturbed morphologies where the inclination was highly uncertain. They are marked with carets in Figure \ref{fig:kinematics}.

We note that the quantity $\sigma_g$ is {\emph {not}} like the typical pressure supported velocity dispersion measured from stellar absorption lines in early-type galaxies. Since $\sigma_g$ is tracing hot gas which can radiate, a high dispersion system can not remain in equilibrium after a crossing time. Therefore $\sigma_g$ is effectively measuring velocity gradients below the seeing limit, as illustrated by the simulations of \citealt{2010ApJ...710..279C}. The typical thermal broadening for T $\sim$ 10$^4$ K Hydrogen gas is 10 km\,s$^{-1}$, so a measure of $\sigma_g$ $>$ 25 km\,s$^{-1}$ is tracing the relative motions of HII regions and/or disordered motions associated with unresolved velocity gradients  (\citealt{2006ApJ...653.1027W}, K12).

\subsection{Quantitative Morphology}

High resolution HST-ACS images allow us to quantify morphologies for each galaxy. In particular, we examine three indices: the Gini coefficient, asymmetry, and concentration. We find little difference between the indices for the V- and I-band and so we simply adopt the I-band measurements (approximately rest V-band at z=0.2) for our analysis. We describe these parameters now and examine how they correlate with kinematics in section 4.

The Gini coefficient was adopted from economic theory. As applied to the study of galaxy morphology, the Gini coefficient provides a non-parametric measurement of the distribution of light among the pixels associated with a galaxy \citep{2003ApJ...588..218A, 2004AJ....128..163L}. The Gini coefficient approaches unity for systems with a large fraction of intensity confined to only a few pixels and approaches zero for a uniform intensity distribution. It compliments the more standard concentration index \citep{1994ApJ...432...75A}, as both trace intensity clustering, but the Gini coefficient is insensitive to the location of the clustering. The typical uncertainty on the Gini coefficient is 0.02, as determined from empirical comparisons of deep (UDF) and shallow fields (GOODS) \citep{2006ApJ...636..592L}.

The concentration index (defined in e.g. \citealt{1994ApJ...432...75A} and \citealt{2002ApJS..142....1S}) measures the ratio of flux contained in two isophotes. The outer isophote is defined by the 2$\sigma$ background contour with a normalized radius of 1 and the inner isophote is defined by a normalized radii of $\alpha$. Concentration, asymmetry, and size were measured with the GIM2D software \citep{2002ApJS..142....1S}, which simultaneously fit to both the V- and I-band images. The concentration index is measured for $\alpha=(0.1, 0.2, 0.3, 0.4)$. We find no significant differences in trends between alpha levels and adopt $\alpha = 0.1$ for our analysis. 

The asymmetry index in GIM2D is as defined in \citet{1994ApJ...432...75A, 1996MNRAS.279L..47A}. Asymmetry is measured by first assigning contiguous pixels to the galaxy, rotating the image by 180$^{\circ}$ and then self-subtracting the mirrored image from the original image.

The error bars we use in the figures below for asymmetry and concentration represent the typical 1$\sigma$ scatter between the four GIM2D measurements for each individual V-Band image. These do not represent the additional uncertainty from the image S/N (limiting magnitude $\sim$ 28.7 AB), for which comparisons with deeper fields are needed. Shallower studies than ours (limiting magnitude $\sim$ 26.5 AB; e.g. \citealt{2015arXiv150401751P}) find relative uncertainties in Concentration and Asymmetry of $\sim$ 0.05 with respect to deeper data (limiting magnitude $\sim$ 27.4 AB).

\section {A Transition Mass in the TFR, the Mass of Disk Formation (M$_{\mathrm{df}}$)}

In the top panel of Figure \ref{fig:kinematics} we show the stellar mass TFR for our sample. There is a noticeable transition at a stellar mass of $\log$ M$_*$/M$_{\odot}\cong$ 9.5. Above this mass, all galaxies in our sample are on the local TFR. Below this mass, galaxies may or may not lie on the local TFR, with the galaxies falling off tending to display systematically high measurements of $\sigma_g$. We posit that $\log$ M$_*$/M$_{\odot}\cong$ 9.5 marks an important stellar mass associated with the stabilization and formation of a disk. We will refer to this mass as ``the mass of disk formation", M$_{\mathrm{df}}$.

In Figure \ref{fig:kinematics} we separate our sample into 3 bins, motivated by their location on the TFR. We define these populations as high mass ($\log$ M$_*$/M$_{\odot} > 9.5$) disks (on the TFR; black triangles), low mass ($\log$ M$_*/M_{\odot} < 9.5$) ordered ($\sigma_g/V_{rot} < 1$) disks (on the TFR; blue circles), and low mass ($\log$ M$_*/M_{\odot} < 9.5$) disordered ($\sigma_g/V_{rot} > 1$) systems (scattering to low V$_{rot}$ from the TFR; red diamonds). This notation will be continued for successive plots.

 We compare the slope in the TFR, $\log$ V$_{rot}$ $\propto$ $\alpha$ $\log$ M$_*$, with a local sample (z $<$ 0.1) of 189 disk galaxies from SDSS \citep{2011MNRAS.417.2347R}, who find a value of $\alpha$ = 0.278 $\pm$ 0.13 over a mass range 9.0 $<$ $\log$ M$_*$/M$_{\odot}$ $<$ 11.0. We refer to this relation as the TF ``ridgeline'' and mark it and its extrapolation to lower stellar masses in our TF plots. The high and low mass disks in our sample are well fit by the local ridgeline.

In the middle panel of Figure \ref{fig:kinematics} we plot  M$_*$ versus $\sigma_g$. As mentioned previously, $\sigma_g$ integrates unresolved velocity gradients in our systems, probing disordered components to the velocity field. The subset of galaxies that fall short of the Tully-Fisher ridgeline (red points) exhibit higher integrated velocity dispersions than the galaxies on the ridgeline at similar stellar mass.

As shown in K07 and K12, combining both the velocity dispersion and rotation velocity into a new kinematic parameter, $S_{K}=\sqrt{KV_{rot}^2+\sigma_{g}^2}$, establishes a relatively tight relation. In the bottom panel of Figure \ref{fig:kinematics} we plot S$_K$ versus M$_*$ for K=0.5. The choice of K=0.5 is motivated by virial arguments for a spherically symmetric tracer distribution (see \citealt{2006ApJ...653.1027W}). Combining both the velocity dispersion and rotation velocity into S$_{0.5}$ re-establishes a tight analogue to the TFR, independent of morphology and coincident with the Faber-Jackson relation (K07, K12).

\begin{figure}
\includegraphics[angle=0,width=8.8cm]{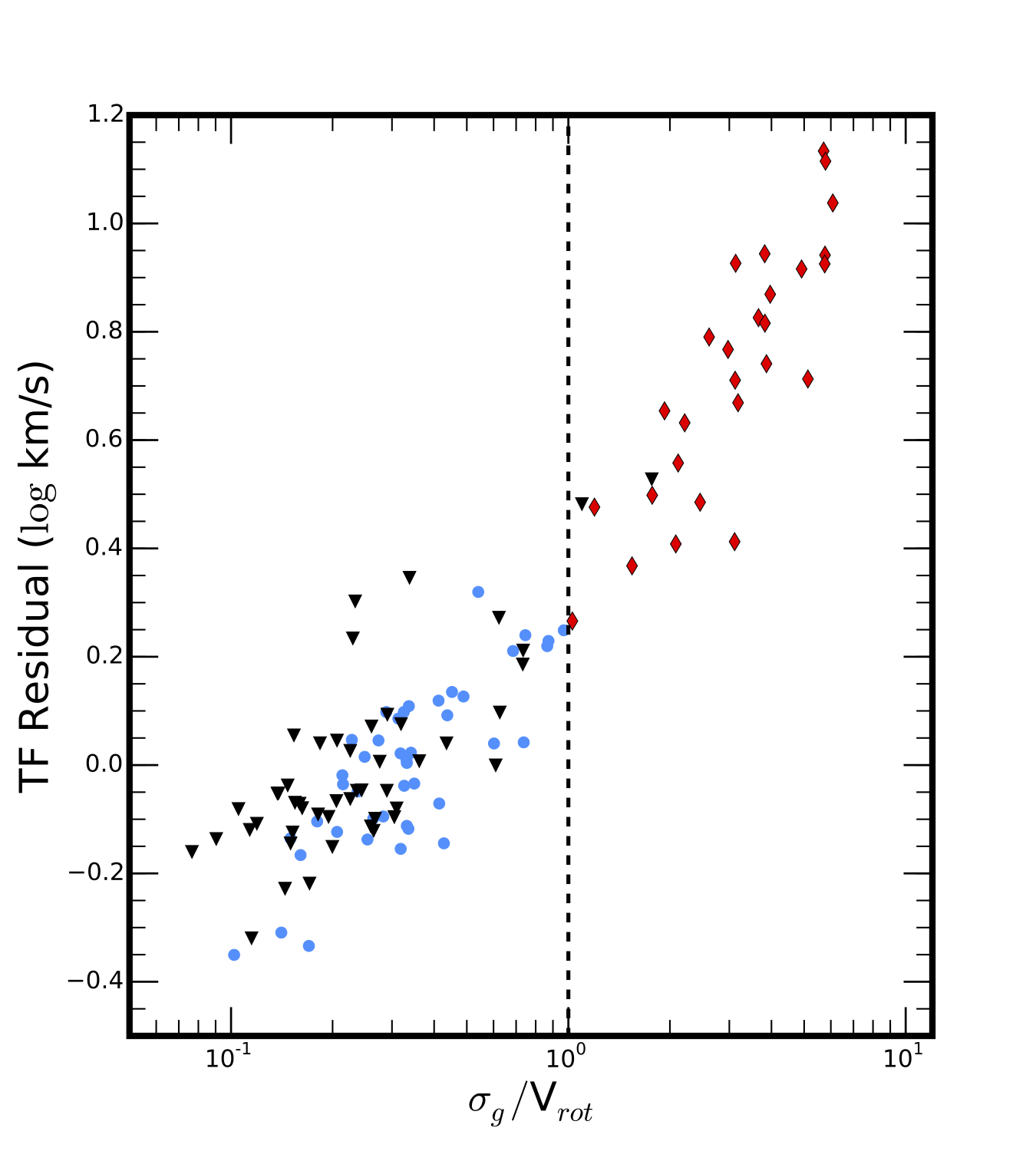}
\caption{The residual from the local Tully-Fisher ridge line \citep{2011MNRAS.417.2347R} is plotted versus the ratio of disordered to ordered motions ($\sigma_g$/V$_{rot}$). As expected from K07 and K12, these quantities trace each other well. We use $\sigma_g$/V$_{rot}$ to quantify the location of a galaxy on the TFR in the remainder of this paper.}
\label{fig:scatter}
\end{figure}

As expected, for galaxies in our sample, the residual from the local TF ridgeline is a strong function of $\sigma_g$/V$_{rot}$. In Figure \ref{fig:scatter} we demonstrate this correlation, demarcating the region where ordered disk galaxies fall (i.e. those on the TFR) would fall. We will use $\sigma_g$/V$_{rot}$ as a proxy for the location of a galaxy on the TFR plot (Figure \ref{fig:kinematics}, top).

\section{Morphology and Gas Kinematics}

\begin{figure*}
\begin{centering}
\includegraphics[angle=0, width=18.4cm]{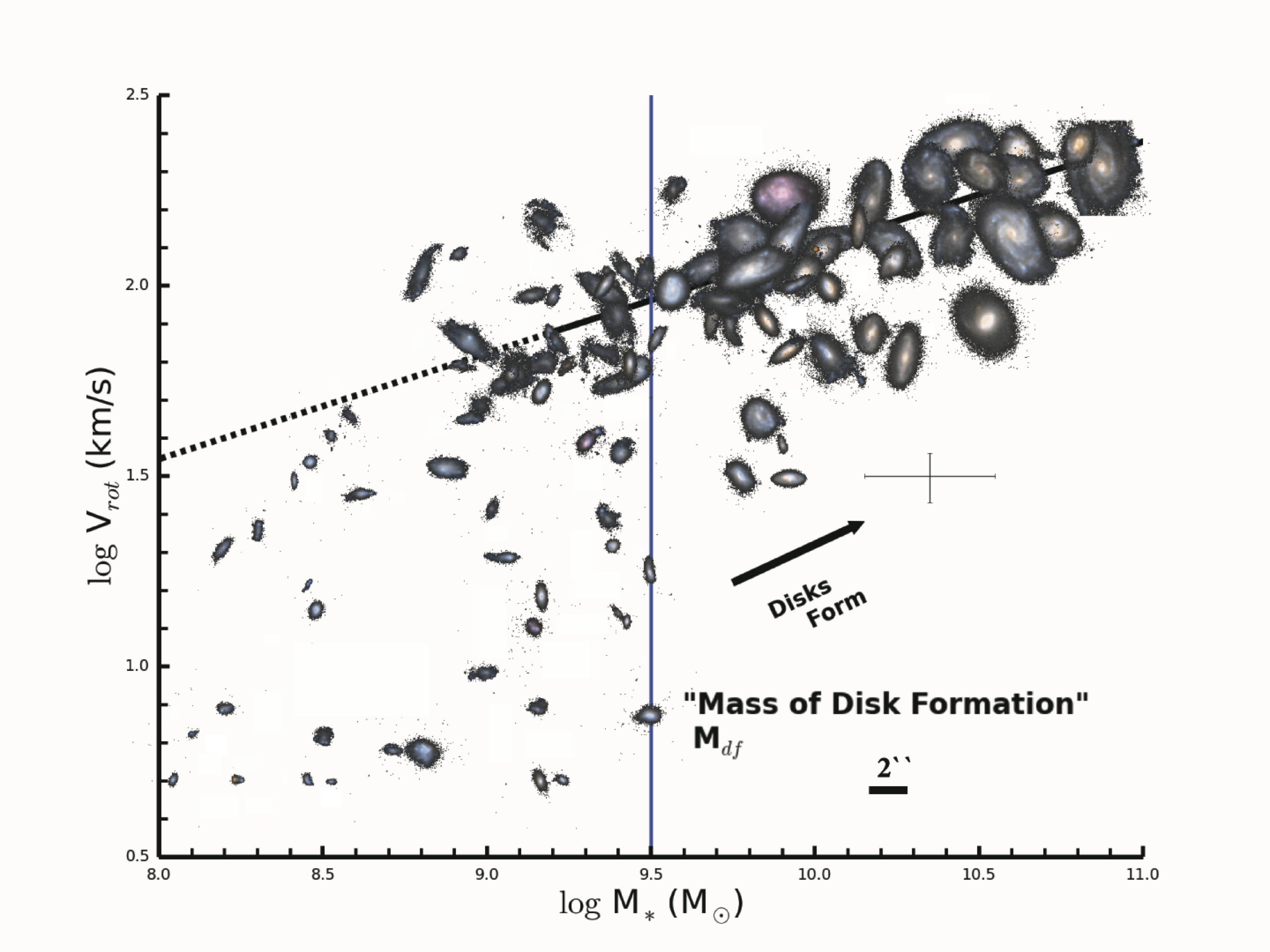}
\caption{The Tully-Fisher relation for a morphologically unbiased sample of blue galaxies over 0.1 $<$ z $<$ 0.375. V and I-band Hubble images are shown in place of points. A 2$\arcsec$ size scale is included for reference. The local TFR from \citet{2011MNRAS.417.2347R} is shown as a solid line and is extrapolated to lower masses with a dotted line. Galaxies with stellar masses $\log$ M$_*$ $\ga$ 9.5 M$_{\odot}$ fall on the TFR and on average have disk-like morphologies. Below this mass, a galaxy may or may not have formed a disk. We therefore call this mass the ``mass of disk formation" (M$_{\mathrm{df}}$). Low mass galaxies which fall from the TFR appear less extended and more irregular than the counterpart galaxies on the TFR. To make the figure manageable, a surface brightness cut on the images was selected based on the higher mass galaxies. This cut leads to a few of the low mass galaxies appearing smaller on this image than their true extent.}
\label{fig:tfmorph}
\end{centering}
\end{figure*}

In Figure \ref{fig:tfmorph} we reproduce the TFR in Figure \ref{fig:kinematics}, but replace the data points with HST V + I-band color images. Upon visual inspection we note distinct morphological characteristics among the regions of this plot identified in Figure \ref{fig:kinematics}. The galaxies with stellar masses higher than $\log$ M$_*/$M$_{\odot} \cong 9.5$ exhibit extended disk-like morphologies. Galaxies with stellar masses $\log$ M$_*/$M$_{\odot} < 9.5$ that are rotation dominated ($\sigma_g$/V$_{rot}$ $<$ 1) also show disk morphology, although they are less extended than the higher mass systems. Galaxies that fall below the TF ridgeline have dispersion dominated kinematics ($\sigma_g$/V$_{rot}$ $>$ 1) and show irregular or compact morphologies.

\begin{figure*}
\includegraphics[angle=0, width=16.3cm]{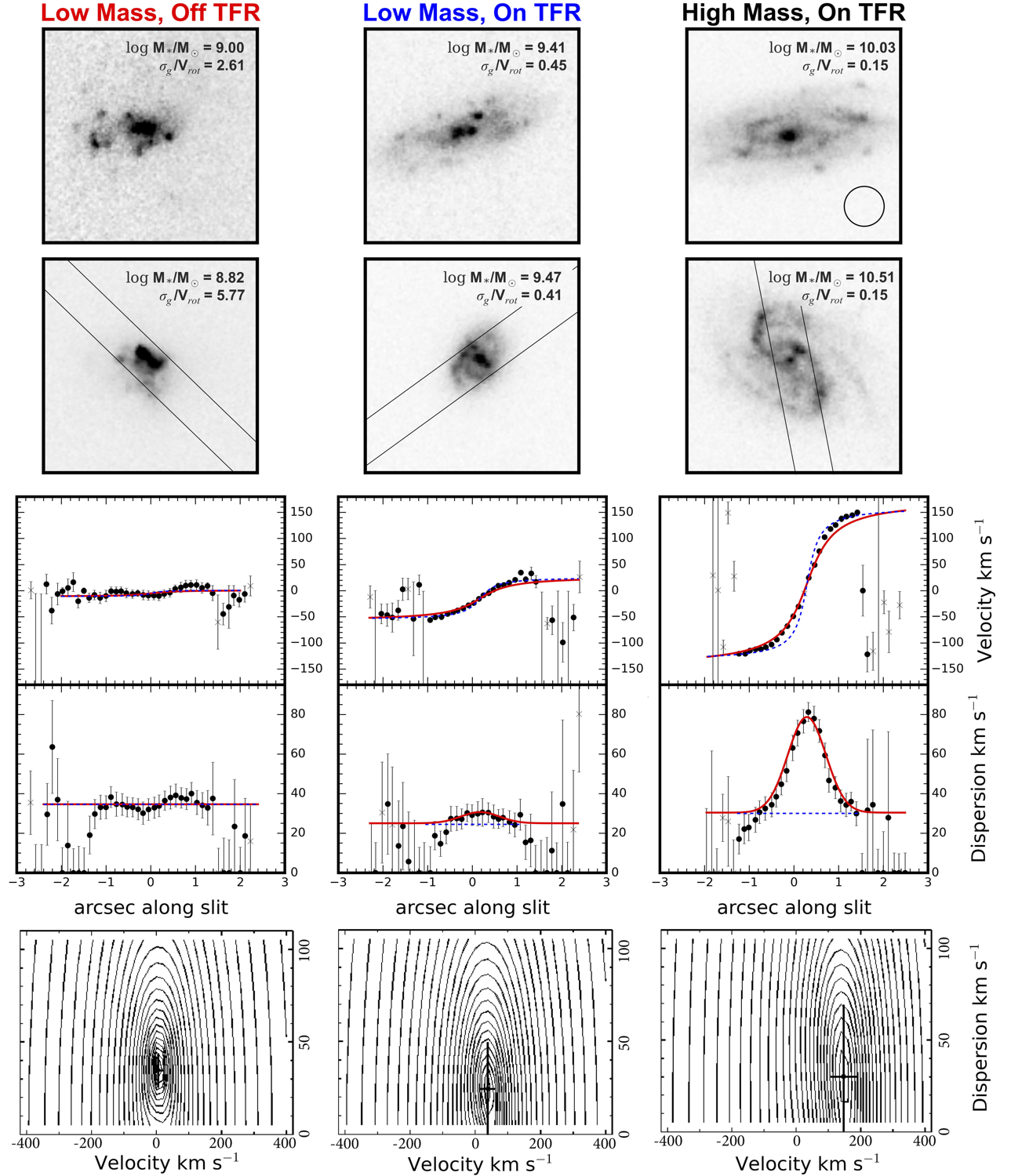}
\caption{Upper Panels: HST V-band images of characteristic galaxies in the three regions of the TFR, Figure \ref{fig:kinematics} (top).  Individual images are 4$\arcsec$ on a side. Outer low surface brightness features are present but not apparent in the images of the low mass images due to the contrast settings chosen. The black circle represents a seeing of 0.75$\arcsec$. Top Left: Low mass galaxies which scatter to low V$_{rot}$ from the TFR are shown. These exhibit compact emission with irregular asymmetric features. Top Middle: Low mass galaxies on the TFR are shown. They exhibit disk-like morphologies characterized by symmetric and elongated intensity distributions. Top Right: High mass ordered galaxies on the TFR are shown. They also exhibit more extended disk-like morphologies. Lower Panels: Example kinematic fits are shown for the three galaxies in the second row. Filled circles were used in the fitting routine, while the crosses were rejected. The blue dashed line is the intrinsic model and the red curve is the seeing blurred model which is fit to the data. For galaxies with rotation gradients, the seeing creates a classic artificial central peak in the velocity dispersion profile and slightly lowers the rotation velocity. Bottom Panels: $\chi^2$ contours for the model parameters.}
\label{fig:indmorph}
\end{figure*}

In Figure \ref{fig:indmorph}, we show two representative galaxies from each of the three defined regions in Figure \ref{fig:kinematics} (from left to right): (1) low mass dispersion dominated galaxies which have compact morphology and irregular asymmetric features, (2) low mass rotationally supported galaxies which have characteristic morphologies of disks: they are more or less symmetric and elongated, and (3) high mass ordered galaxies which have similar disk-like morphologies.

These trends were noted by visual inspection in previous work (K07). We will now demonstrate that gas phase kinematics are related to quantitative galaxy morphology by comparing $\sigma_g$/V$_{rot}$ (ratio of disordered to ordered motions) to quantitative morphological indices (Gini coefficient, concentration, asymmetry). As described in the rest of this section, we find that the galaxies which fall on the Tully-Fisher ridgeline tend to exhibit low values of asymmetry, concentration, and Gini. In addition, they tend to have larger sizes consistent with the disk size-mass relation.  Kinematically ordered galaxies ($\sigma_g$/V$_{rot}$ $<$ 1) at both low and high mass are consistent with being drawn from identical parent distributions for these indices. Galaxies which are kinematically disordered ($\sigma_g$/ V$_{rot}$ $>$ 1) and fall from the TFR ridgeline tend to exhibit quantitative morphologies characteristic of disturbed or compact systems (high asymmetry, concentration and Gini) and are statistically distinct from the galaxies on the TF ridgeline. Below we detail these findings.

\begin{figure}
\includegraphics[angle=0,scale=.35]{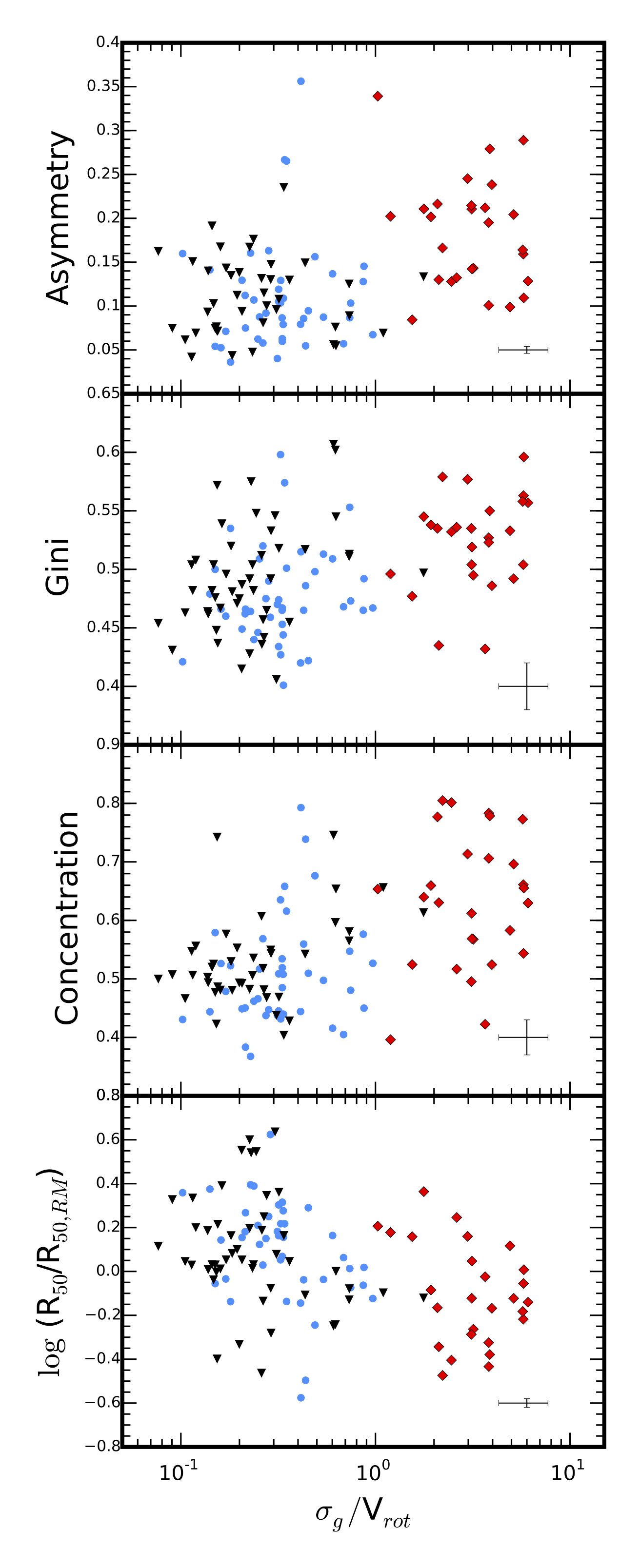}
\caption{Quantitative morphological indices and V-band half-light radius (R$_{50}$) are shown as a function of kinematics for the galaxies in our sample. The bottom panel shows the log residuals from the SDSS g+r derived disk mass-size relation (RM; \citealt{2011MNRAS.410.1660D}). The quantity $\sigma_g$/V$_{rot}$ measures the relative contributions from disordered motions and ordered rotation. Galaxies which scatter to low V$_{rot}$ off of the TF ridgeline (red diamonds) are on average more asymmetric, have higher Gini and concentration values, and are smaller than galaxies on the TF ridgeline (blue circles for low mass and black triangles for high mass).}
\label{fig:morph}
\end{figure}

\begin{figure}
\includegraphics[angle=0,scale=0.40]{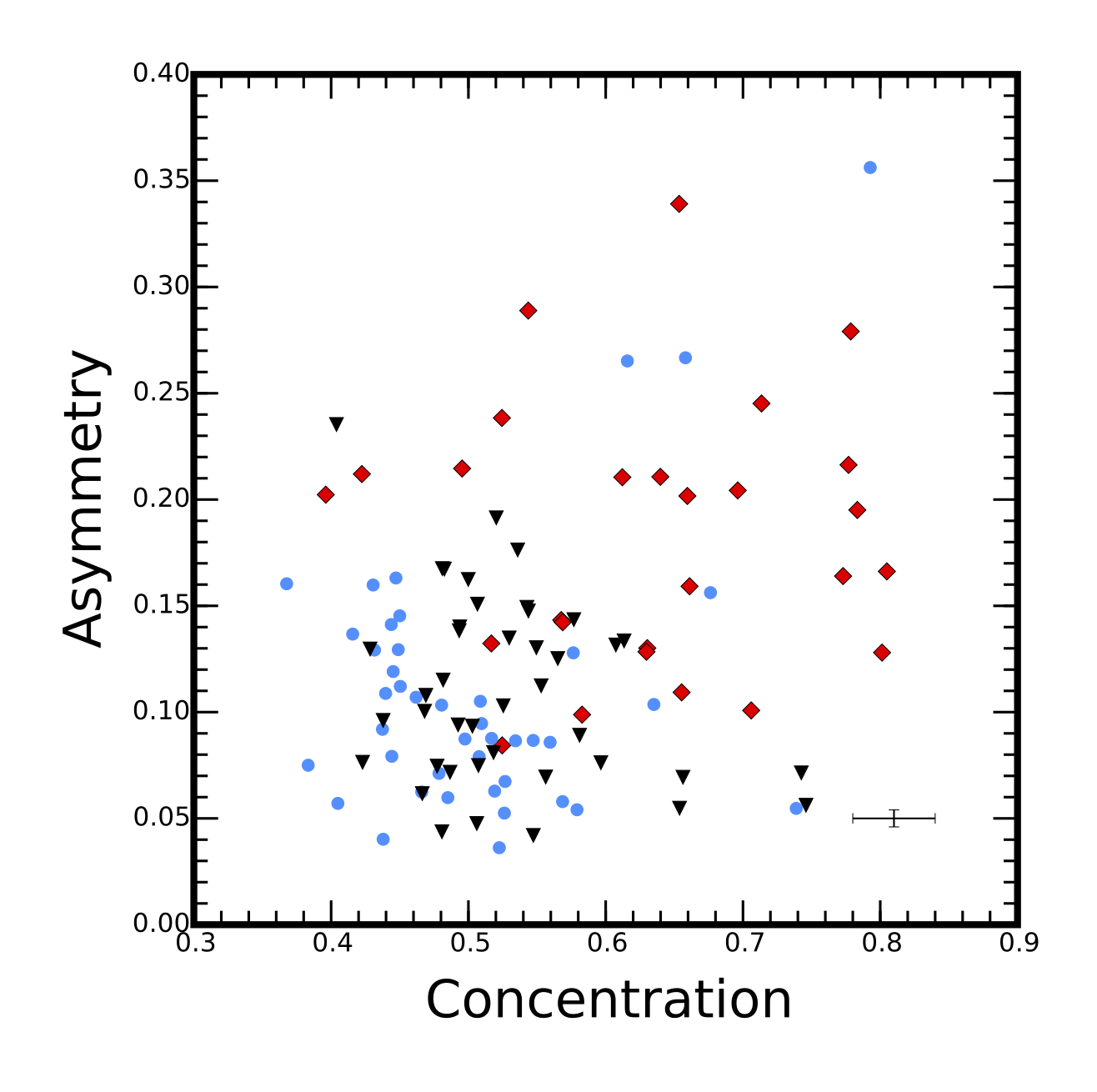}
\caption{In the concentration-asymmetry plane, the high and low mass systems (black triangles and blue circles, respectively) are relatively localized to low asymmetry and concentration values. Low mass dispersion dominated galaxies (red diamonds) tend to scatter to larger values of both asymmetry and concentration.}
\label{fig:conc-asymm}
\end{figure}

The top panel in Figure \ref{fig:morph} shows the asymmetry index versus $\sigma_g$/V$_{rot}$ for our galaxy sample. Low mass disordered galaxies (red diamonds) on average have a higher distribution in asymmetry than the galaxies with ordered kinematics which lie on the TF ridgeline (blue circles and black triangles). The asymmetry index is distributed with an average and standard deviation 0.18 $\pm$ 0.06 for low mass disordered systems and 0.11 $\pm$ 0.06 and 0.11 $\pm$ 0.04 for the low and high mass ordered systems, respectively. We run a two-sample Kolmogorov-Smirnov (K-S) test to determine the probability that the three distributions in asymmetry are drawn from identical parent distributions. The p-value returned from the K-S test represents the probability that data drawn from identical parent distributions would be as disparate as observed. The K-S test between the low and high mass ordered populations yields a p value of 0.77, indicating a high probability of being drawn from identical parent distributions. However, the K-S tests between the low mass disordered galaxies and the high/low mass ordered distributions yield near zero probabilities of being identical (p = 10$^{-5}$ and 10$^{-4}$, for comparisons with high and low mass ordered samples, respectively.)

The second panel in Figure \ref{fig:morph} shows the Gini coefficient versus $\sigma_g$/V$_{rot}$. The disordered low mass population of galaxies exhibits a trend towards higher values of Gini. Average values of Gini increase from G = 0.48 $\pm$ 0.04 and 0.48 $\pm$ 0.09 for the low and high mass ordered systems, respectively, to G = 0.52 $\pm$ 0.04 for the disordered low mass systems. The K-S test between the ordered populations yields a p value of 0.48, while the dispersion dominated systems are unique compared with either distribution of ordered galaxies (p = $10^{-5}$). While statistically significant, the relative change in Gini is small, reflecting the fact that even disk galaxies exhibit a level of non-uniformity (i.e. spiral features and clumpy star-formation). In large heterogeneous samples of galaxies the range of the Gini coefficient is relatively small (e.g. \citealt{2008ApJ...672..177L, 2015arXiv150401751P}), with a floor at a value of around 0.40 for the most uniform galaxies and a peak value of only 0.65. In relation to this small range, a change of 0.04 reflects a non-negligible difference in galaxy type.

The third panel of Figure \ref{fig:morph} shows the concentration index versus $\sigma_g$/V$_{rot}$. Similar to the Gini coefficient, kinematically disordered systems typically display higher values of concentration.  The concentration index increases from 0.51 $\pm$ 0.09 and 0.53 $\pm$ 0.07 for the low and high mass ordered systems, respectively, to 0.65 $\pm$ 0.12 for the disordered low mass systems. The K-S test between the ordered populations yields a p value of only 0.02, due to the large scatter in the low mass ordered distribution. The p value between the low mass disordered distribution and the ordered distributions is once again near zero (p = 10$^{-5}$).

We now examine the correlation between size with disordered motions in our sample. We use the HST V-band half-light radius R$_{50}$ as measured using GIM2D. The low mass dispersion dominated galaxies have a mean value of R$_{50}$/kpc = 1.8 $\pm$ 1.2 rms. The low and high mass ordered systems are close to a factor of 2 larger, 3.1 $\pm$ 1.5 and 4.5 $\pm$ 2.0 kpc, respectively. To account for the well known mass dependance on size, we examine residuals from the SDSS DR7 median fit to the disk mass-size relation (Equation 3 in \citealt{2011MNRAS.410.1660D}).  In the bottom panel of Figure \ref{fig:morph} we plot the log residual from the size mass relation ($\log ($R$_{50}/$R$_{50,RM})$) versus $\sigma_g$/V$_{rot}$. The ordered galaxies in our sample are distributed around the relation with an average scatter of 0.22 dex, consistent with the intrinsic width of the relation. The residuals for the dispersion dominated systems are centered 0.15 dex below the relation, although there are a few outliers that lie on or above the relation. At a given mass, the dispersion dominated systems tend to be smaller than expected from the radius-mass relation.

In the concentration-asymmetry plane (Figure \ref{fig:conc-asymm}) these differences become more apparent, with both the low and high mass ordered galaxies occupying relatively narrow regions of low asymmetry and concentration. The dispersion dominated systems, however, tend to exhibit markedly high scatter in this plane, tending towards higher values of both concentration and asymmetry.

\emph{In summary: The ordered systems in our sample display quantitative indices characteristic of disks: low asymmetry and low concentration. Kinematically disordered galaxies on the other hand exhibit concentrated emission with asymmetric features, statistically distinct from their ordered counterparts.}

\section{Comparison with Previous Studies}

\subsection{Local Universe}

Large local studies of the TFR have primarily focused on massive disk galaxies with $\log$ M$_*$/M$_{\odot}$  $\ga$ 9.5 (e.g. \citealt{2001ApJ...563..694V, 2007AJ....134..945P, 2007ApJ...671..203C, 2010ApJ...716..198B, 2011MNRAS.417.2347R}). We find that the majority of star-forming galaxies above this mass are morphologically disk-like and the TFR is well behaved. In fact, the only populations of dispersion dominated systems at high stellar masses in the local Universe are the relatively rare and interaction induced (U)LIRGs systems (e.g. \citealt{2014A&A...568A..14A}) and merging galaxies (e.g. the HI Rogues Gallery of \citealt{2001ASPC..240..657H}).

The TFR for $\log$ M$_*$/M$_{\odot}$ $<$ 9.5 is less well-studied, especially for a morphologically unbiased sample. Furthermore, velocity dispersion is not frequently measured. Integrated line widths are often used as a proxy for the rotation velocity, with corrections assumed for anomalous contributions from random motions.

Most TFR studies of dwarf galaxies have focused on gas dominated systems (M$_g$ $\geq$ M$_*$). These galaxies typically show large residuals from the TFR to lower masses at a given rotation velocity, in the opposite direction to our sample. A tight baryonic Tully-Fisher relation is established once gas masses are included \citep{2005ApJ...632..859M} and it has been well calibrated for a local sample of gas-dominated galaxies (e.g. \citealt{2009AJ....138..392S}).

For local dwarf galaxies with both $\sigma_g$ and V$_{rot}$ measured, evidence for a luminosity dependence on V/$\sigma_g$ has been long established (e.g. \citealt{1998ARA&A..36..435M}). This trend is to be expected even for galaxies on the TFR, if we assume that $\sigma_g$ is independent of mass and V$_{rot}$ declines with mass. When available, though, resolved kinematics of irregular or compact dwarf systems typically show complex velocity fields which depart from a regularly rotating thin disk (e.g. \citealt{2001AJ....121..625B, 2002AJ....123.2358K, 2005AJ....130.1593V}). For instance, the well studied SMC (M$_*$/M$_{\odot}\sim$ 10$^{8.7}$) and LMC (M$_*$/M$_{\odot}\sim$ 10$^{9}$) are both rotationally supported in their stellar and neutral gaseous components (V/$\sigma\sim$ 3), while the kinematics of the smallest local dI systems, traced through neutral HI gas, typically showing no signs of rotation (e.g. \citealt{2003ApJ...592..111Y}).

Recent surveys of HI in the local volume have helped further characterize the dynamics of dwarfs, e.g. THINGS \citep{2008AJ....136.2563W}, FIGGS \citep{2008MNRAS.386.1667B}, VLA-ANGST \citep{2012AJ....144..123O} and LITTLE THINGS \citep{2012AJ....144..134H}. These surveys find mixed evidence among dwarf galaxies for both the existence of (i) rotationally supported disks (e.g. \citealt{2008MNRAS.386..138B, 2008AJ....136.2563W}) and (ii) systems with thick disks or high/dominant contributions of dispersion (e.g. \citealt{2010MNRAS.404L..60R, 2012AJ....144..123O}).

The current data for resolved ionized gas kinematics in low mass galaxies is more limited. Interest in actively star-forming blue compact dwarfs (BCD) has led to measurements of velocity fields for a relatively large sample ($\sim 100$) of these systems, usually in HI (e.g. \citealt{1998AJ....116.1186V, 2001AJ....122..121V, 2004AJ....128..617T}). Star-bursting systems, of which BCDs are a subset, make up close to 5$\%$ of the population of local dwarf galaxies \citep{2009ApJ...692.1305L}. The origin of the compactness and late stage star-formation in the low mass ($\log$ M$_*$/M$_{\odot}$ $\sim$ 7.0 - 9.6; \citealt{2013ApJ...764...44Z}) BCDs is up for debate. Recent HI investigations suggest high fractions of kinematically disturbed disks ($\sim$ 50\%), although only a handful of systems fall short of the baryonic Tully-Fisher relation \citep{2014A&A...566A..71L}. The ionized velocity fields are often found to exhibit high or dominant contributions from dispersion as well (e.g. \citealt{2011MNRAS.418.2350P}).

One of the largest surveys of resolved ionized gas kinematics in local galaxies is the GHASP survey (Gassendi H$\alpha$ Survey of Spirals; \citealt{2008MNRAS.390..466E, 2010MNRAS.401.2113E}). The GHASP survey selection was focused on galaxies with both spiral and irregular morphologies, although most of the more luminous galaxies display characteristic disk morphologies. The GHASP survey samples a wide range in stellar mass 9.0 $<$ $\log$ M$_*$/M$_{\odot}$ $<$ 11.7 \citep{2008MNRAS.390..466E}. The TFR for a subset of galaxies is found to be relatively tight down to M$_K$ = -18 or $\log$ M$_*$= 8.0 M$_{\odot}$ \citep {2011MNRAS.416.1936T}. While galaxies with large non-circular motions and corresponding rotation curve asymmetries show the largest scatter from the TF relation, to both low V$_{rot}$ and low luminosity, the scatter is small compared to the dispersion dominated systems in our sample. Aside from a few outliers, no large break to low rotation velocity was found in the TFR for the GHASP sample. This may be due to a combination of both small sample sizes at masses $\log$ M$_*$/M$_{\odot}$ $<$ 9.5 (9 galaxies) and/or selection functions for rotating galaxies playing a role in constructing the stellar mass TFR.  The $\sigma$/V for these 9 low mass galaxies ranges between 0.1 - 0.7 (i.e. all would be kinematically ordered in this paper). On visual inspection of their optical morphologies in NED, we note that seven of these galaxies appear to have disk morphologies, with the presence of spiral arms in most cases. The low $\sigma$/V in these galaxies is consistent with the picture presented in this paper, that disk morphologies indicate ordered kinematics. Two of the galaxies (UGC 5721, UGC 10757)  do appear to have disturbed morphologies, so it is interesting to note that their $\sigma$/V is still less than 1 (0.16 and 0.35, respectively). The mean velocity dispersion for the GHASP sample is 25 $\pm$ 5 km s$^{-1}$, typical of local disks in general. This level of dispersion is expected for a combination of the thermal broadening from H atoms (T$\sim$10$^4$ K; $\sim$ 10 km\,s$^{-1}$) and internal turbulence of HII regions ($\sim$ 20 km\,s$^{-1}$; \citealt{1990ARA&A..28..525S}). These measurements are comparable to the mean velocity dispersions for our ordered disks (21.0 $\pm$ 8.1 km\,s$^{-1}$) and slightly lower than our systems with more disturbed kinematics (34 $\pm$ 7.3 km\,s$^{-1}$). For the few compact rotators in GHASP we ran mock tests (see Appendix A for an example with galaxy UGC 528) to ensure that we could accurately recover their rotation if they were in our sample at z $\sim$ 0.2. The results of these tests indicate that if these galaxies were included in our sample we would recover their rotating disks.

Our understanding of the ionized kinematics of dwarf galaxies at low redshifts will substantially increase in the near future with ongoing and upcoming Integral Field Spectroscopy (IFS) surveys. The CALIFA (0.005 $<$ z $<$ 0.03, R $\sim$ 850-1650; N $\sim$ 600), SAMI (z $<$ 0.12, R $\sim$ 1700-4500; N $\sim$ 3400) and MaNGA (z $\sim$ 0.03; R $\sim$ 2000; N $\sim$ 10,000) surveys will provide velocity fields for large ensembles of local low mass galaxies. First results from CALIFA \citep{2015A&A...573A..59G} have reported that about half of the galaxies in their sample (8.5 $<$ $\log$ M$_*$/M$_{\odot}$ $<$ 11.5)  have structure in their velocity fields which departs from a rotating disk. However, resolved velocity dispersions were not measured in this study and the relatively coarse spectral resolution (R $\sim$ 1200) may not allow for measurements of the dispersion at the differences outlined in this paper. First results from SAMI show a large fraction of both low and high mass local galaxies falling to low V$_{rot}$ from the TFR \citep{2014ApJ...795L..37C}. The SAMI survey improves upon the GHASP sample with a large coverage of low mass systems, pushing down to stellar masses $\log$ M$_*$/M$_{\odot}$ $\sim$ 8. Below a stellar mass $\log$ M$_*$/M$_{\odot}\cong$ 9, they find a precipitous trailing of galaxies from the well-defined 1 R$_e$ TF relation \citep{2007MNRAS.377..507Y}. High spectral resolution in the red band (6300-7400 $\AA$, R $\sim$ 4500) allow measurements of velocity dispersions. A tight scaling relation (M$_*$-S$_{K}$) is established once they account for the average measured velocity dispersion, similar to previous findings at high redshift \citep{2006ApJ...653.1049W, 2007ApJ...660L..35K, 2010A&A...510A..68P}. The SDSS-III MaNGA survey will provide extensive additional coverage of this low mass regime, with a flat selection distribution down to a stellar mass of 10$^9$ M$_{\odot}$ and sampling out to at least 1.5 times the effective radius (R$_e$) \citep{2015ApJ...798....7B}. 

In summary, when studied in large numbers, local studies of the TFR  have been biased towards massive disk galaxies and moreover will often not have the spectral resolution to resolve velocity dispersion. When resolved gas phase kinematics are available in low mass galaxies, complex velocity fields are often found in both neutral and ionized gas. The local TFR for dwarf galaxies will be expanded upon in the upcoming years, with surveys probing kinematics in ionized gas to low stellar masses in unprecedented numbers. Early results indicate an increase in complexity in the resolved 2D velocity fields of these low mass galaxies.

\subsection{Low-Intermediate Redshift}

Due primarily to sensitivity limits, there have been few studies of the kinematics in more distant (z $>$ 0.1) dwarf galaxies. Furthermore, most studies do not measure resolved velocity dispersions and do not include irregular or compact galaxies.

Using deep 8 hour DEIMOS exposures, \cite{2014ApJ...782..115M} constructed the stellar mass TFR for half of their sample (N = 41) of dwarf galaxies (7 $< \log$ M$_*/$M$_{\odot} <$ 9) out to z = 1, demonstrating that at least a fraction of dwarf galaxies are settled into rotationally supported systems at intermediate redshift. The remaining galaxies were either missing emission or were too small to reliably resolve rotation.

Recently, interest in dispersion dominated galaxies at high redshift (e.g. \citealt{2009ApJ...697.2057L, 2009ApJ...706.1364F}) has motivated searches for high dispersion low redshift analogs. These studies typically select highly star-forming systems and find complex kinematics with low V/$\sigma_g$ compared to local disk galaxies (e.g. \citealt{2010ApJ...724.1373G, 2012ApJ...754L..22A, 2014MNRAS.437.1070G}). 

Compact Lyman Break analogues (LBA) are an example of such a selection. LBAs are highly star-forming galaxies at z $\sim$ 0.2, selected from their high GALEX NUV luminosity \citep{2005ApJ...619L..35H}. Although these systems have slightly higher stellar masses 9.5 $\le$ M$_*$/M$_{\odot}$ $\le$ 10.7 than the break in our sample, they exhibit morphologies and kinematics which are similar to some of our more compact dwarf galaxies. \citet{2010ApJ...724.1373G} report on AO-assisted IFS observations of 19 compact LBAs at z $\sim$ 0.2 with stellar masses near our transition mass, 9.1 $<$ $\log$ M$_*$/M$_{\odot}$ $<$ 10.7. They find a very high contribution of disordered motion in these rare systems ($\sigma_g$ $\sim$ 70 km s$^{-1}$) and evidence that disk-like structure is more common in high mass LBAs. Similarly, \citet{2014MNRAS.437.1070G} measured H$\alpha$ kinematics for 67 extremely star-forming systems down to a stellar mass of $10^9$ M$_{\odot}$ and found that one fifth of their sample shows no signs of rotation. The rotating disk galaxies in their sample fall on the TFR while still exhibiting values of V/$\sigma_g$ lower than local disks,  comparable to dynamically hot high redshift galaxies. The dispersion dominated galaxies V/$\sigma_g$ $\le$ 1 in both the intermediate redshift IFS samples have typical stellar masses above our observed mass of disk formation ($\log$ M$_{\mathrm{df}}$/M$_{\odot}$ $\sim$ 9.5). Although the unique selections (e.g. high star-formation, high surface mass density) for these samples are not necessarily representative, they do demonstrate the important contribution of $\sigma_g$ to the kinematics in active systems.

At high redshifts (1.5 $<$ z $<$ 3), complex velocity fields and thick disks with high velocity dispersions are ubiquitous in the samples currently available in the literature (e.g. \citealt{2009ApJ...697.2057L, 2009ApJ...706.1364F, 2015ApJ...799..209W}). A stellar mass dependence on the relative contributions of $\sigma_g$ and V$_{rot}$ appears to exist, although beam smearing (see Appendix) may provide the enhanced dispersion in some of the the smallest systems \citep{2013ApJ...767..104N}. Kinematic data for low mass galaxies at these redshifts is currently limited. The advent of high sensitivity near-IR instruments (e.g. MOSFIRE, KMOS) will soon provide large samples of kinematics for galaxies with low stellar masses.

\section{Conclusions}

We study the stellar mass Tully-Fisher relation (TFR; stellar mass versus rotation velocity) for a morphologically blind sample of
emission line galaxies in the field.  The galaxies are at a redshift of $\sim0.2$, or a lookback time of about 2 Gyrs, and therefore
we expect little evolution from the local TFR.  

We report on a transition mass in the TFR which we call the 
``mass of disk formation," M$_{\mathrm{df}}$.  This mass separates galaxies which always form disks (masses greater than M$_{\mathrm{df}}$)
from those which may or may not form disks (masses less than M$_{\mathrm{df}}$).
For M$_* > M_{df}$, all galaxies in our sample are settled onto the local TFR.  However, for  M$_* < M_{df}$, galaxies can either
lie on the TFR or scatter off of it to low rotation velocity. The galaxies which scatter off have higher disordered motions, as measured through integrated gas velocity dispersions ($\sigma_g$). Moreover, we find that galaxies on the TFR are morphologically distinct from those which scatter off.  The quantitative morphologies of galaxies on the relation are on average less asymmetric and
concentrated than those galaxies which scatter off.  

We perform mock observations of the 3D kinematics of local galaxies to investigate how well we are able to recover the rotation velocity of the smallest galaxies in our sample.  These simulations show that we are able to recover rotation, if present, in the smallest galaxies in our sample.

\section*{Acknowledgements}

The authors would like to thank the anonymous referee who provided useful suggestions that improved this manuscript. RCS gratefully acknowledges support through a grant from the STScI JDF. This research has made use of the Fabry Perot database, operated at CeSAM/LAM, Marseille, France. The authors also acknowledge NSF grants AST 95- 29098 and 00- 71198 to UC Santa Cruz. RCS would like to thank A. Dutton for helpful comments. We wish to extend thanks to those of Hawaiian ancestry on whose sacred mountain we are privileged guests.

{}

\appendix

\renewcommand{\thefigure}{A\arabic{figure}}
\begin{figure*}
\begin{centering}
\includegraphics[width=1.0\textwidth]{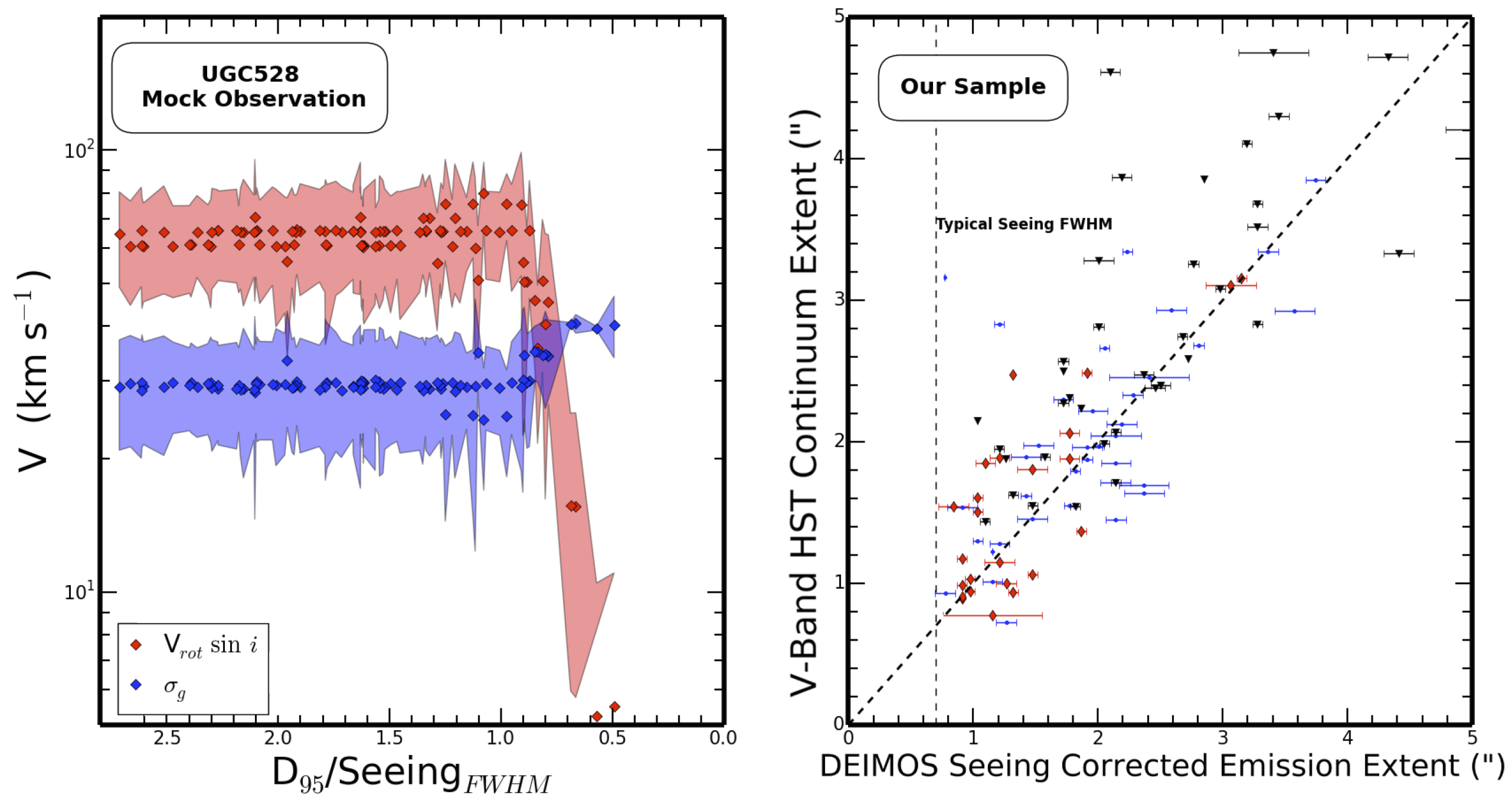}
\caption{Left: Mock observations of UGC 528 at varying apparent sizes. The seeing is kept constant at 0.75$\arcsec$ for all mock observations. We are able to recover the true rotation velocity and average gas velocity dispersion down to a diameter, D$_{95}$ = (0.87 $\pm$ 0.06) $\times$ seeing. Right: Measurements of the continuum and emission line extent for the galaxies in our sample. The continuum extent is defined as 4 $\times$ $\sigma_{cont}$ of the intensity profile. For a Gaussian profile, this will equal D$_{95}$. For galaxy profiles with broader wings than a Gaussian, D$_{95}$ will be larger than this value. The continuum is derived from the V-band image (rest B at z = 0.2) and should trace the young star forming regions. The seeing-corrected DEIMOS emission line extent is defined as 4 $\times$ $\sigma_{emission}$.  The seeing for the observations in our sample ranged between 0.55$\arcsec$-1.20$\arcsec$ The number of galaxies for each bin in seeing are: [0.55 - 0.70$"$] 26 galaxies, [0.70 - 0.85$"$] 44 galaxies, [0.85 - 1.0$"$] 35 galaxies, [1.0 - 1.2$"$] 14 galaxies. The typical seeing for our sample is demarcated by the vertical black line. All of the galaxies in our sample are sufficiently large enough to reliably measure a rotation velocity, if present.}
\label{fig:ugc528}
\end{centering}
\end{figure*}

\section{Modeling the effects of apparent size on measuring kinematics}
Reliably measuring resolved kinematics for a galaxy depends critically on its apparent size relative to the seeing, i.e. beam smearing (\citealt{1987PhDT.......199B}). Atmospheric turbulence blurs together intrinsically spatially separate velocity gradients, leading to a smoothing of the rotation field and a boosted central velocity dispersion (e.g. \citealt{2006ApJ...653.1027W}). ROTCURVE models the seeing and recovers rotation curves and average integrated gas velocity dispersions for galaxies with apparent sizes larger than the seeing (see examples in \citealt{2006ApJ...653.1027W}). Overcoming beam smearing with galaxies whose intrinsic size is comparable to the seeing is challenging and has been addressed by several papers with respect to IFS measurements at high redshift (e.g. \citealt{2010MNRAS.401.2113E, 2011ApJ...741...69D}). 

To understand the effects of beam smearing on our measurements we perform mock observations of observed velocity and dispersion maps for a nearby galaxy from the literature. In particular, we use Fabry-Perot observations of UGC 528 from the GHASP survey \citep{2010MNRAS.401.2113E}. We determine the apparent galaxy size (defined as the extent of the line emission) at which the intrinsic rotation velocity is no longer recovered under the observing conditions and instrumental set-up used by the DEEP2 Survey.

\subsection{Mock Observations of UGC 528}

UGC 528 is a small rotating local spiral galaxy of Hubble Type SAB(rs)b \citep{2003A&A...399...51G}.  It has a rotation velocity V$_{rot}\sin$(i) $\sim$ 60 km\,s$^{-1}$ and a relatively high average dispersion across its face of $\sigma_g$ = 26.0 $\pm$ 9.6 km\,s$^{-1}$. 

We build mock spectral cubes (3D: spatial $\times$ spatial $\times$ $\lambda$) from the available velocity, dispersion and flux maps. This allows us to perform mock observations for an arbitrary velocity sampling without needing to interpolate between slices. To do this, we first construct an empty cube with velocity sampling matching DEEP2, namely $\Delta$v = 12.6 km\,s$^{-1}$ at z = 0.2. Next, using the flux, velocity and dispersion maps we fill the cube with model H$\alpha$ emission. For each spaxel in the map with valid velocity information we create a Gaussian line profile with a mean line of sight velocity V$_{los}$, dispersion $\sigma_g$ and amplitude scaled to match the integrated flux. This cube is then projected from its original redshift to z = 0.2 by resizing the spaxel scale. Each slice in the cube is then convolved with a 2D spatial Gaussian kernel (FWHM = 0.75$\arcsec$) to simulate seeing.

A mock 1$\arcsec$ wide slit is placed along the kinematic PA of the galaxy. From this we produce a mock spectrum using the DEEP2 pixel scales (0.118 $\arcsec$/pixel, 0.33 $\AA$/pixel) and spectral resolution (R $\sim$ 5000). Noise is added to the 2D spectrum to match the typical pixel to pixel S/N profiles of our faintest sources. 

We create separate cubes in this manner, each time varying the spatial size to simulate different apparent sizes for UGC 528. The final product is a set of 100 mock observations of UGC 528 at z = 0.2 ranging in size 0.4 -2.2$\arcsec$ with a 0.018$\arcsec$ step size. Each mock observation results in a 2D spectrum of the emission line for a given apparent size.

Next we run each of the mock spectra through ROTCURVE to measure kinematics. We fix the scale radius of UGC 528 in physical units to its true value of 0.6 kpc \citep{2010MNRAS.401.2113E}. For reasonable values, we find that the chosen scale radius has little impact on the kinematic measurements, as was also found by \citet{2006ApJ...653.1027W}. We measure V$_{rot}$ $\times$ $\sin(i)$ and $\sigma_g$ for each of the 100 mock spectra. The results of our test are demonstrated in Figure \ref{fig:ugc528}. As expected, at large intrinsic sizes relative to the seeing  (D$_{95}$ $>$ 1.5 $\times$ seeing) we are able to recover the intrinsic rotation and integrated dispersion well. As the mock observations reach closer to and beyond the scale of the seeing, we note that the measured rotation velocity decreases and the measured dispersion increases as expected. To quantify the scale at which this turnover occurs, we fit a simple change point model using PyMC \citep{Patilpymc}. The turnover to artificially low V$_{rot}$ occurs at a scale of D$_{95}$/Seeing = 0.87 $\pm$ 0.06. This error incorporates both the measurement error imposed by the slit (red shading in Figure \ref{fig:ugc528}, left) and the error in the fit to the model.

{\emph {We conclude from these mock observations that we can reliably measure rotation velocities for the galaxies in our sample down to D$_{95}$/Seeing = 0.87 $\pm$ 0.06}.}

\subsection{Emission sizes vs HST Continuum sizes}
We now compare the intrinsic sizes of the galaxies in our sample to the size of the seeing. In the right panel of Figure \ref{fig:ugc528} we plot the size of the galaxy in emission, tracing areas of star-formation, against the size in continuum, tracing the young stellar population. The Hubble sizes are measured from the HST V-Band images and the emission sizes are measured directly from the DEIMOS spectra. The observed emission lines ($\sigma_{em,obs}$) are broadened by the combined PSF of the instrument and the seeing ($\sigma_{PSF}$). We can recover the true extent of the emission line ($\sigma_{em,corr}$): 

\begin{equation}
\sigma_{em,corr}=\sqrt{\sigma_{em,obs}^2-\sigma_{PSF}^2}
\end{equation}

For an approximate 1-D Gaussian profile, the diameter containing 95$\%$ of the flux will be 4 times this value. In reality, intrinsic emission profiles of galaxies will be broader than a Gaussian distribution and the value of $\sigma_{em,corr}$ is primarily determined by the profile of the core. A galaxy profile with broader wings than a Gaussian will have a D$_{95}$ that is larger than that given by 4 $\times$ $\sigma_{em,corr}$. The HST V-band (rest B at z $\sim$ 0.2) extent correlates well with the seeing-corrected emission line sizes (Figure \ref{fig:ugc528}, right). We demarcate the typical seeing size of 0.75$\arcsec$.

The smallest galaxies in our sample reach down to seeing-corrected emission extents of $\sim$ 1.2$\arcsec$ (or 1.6 times the typical seeing FWHM), well into the predicted regime for accurately recovering the kinematics (Figure \ref{fig:ugc528}; left panel). In conclusion, it is unlikely that our low mass dispersion dominated galaxies are dispersion-dominated because of beam-smearing.

\end{document}